# Applications of Quantum Randomness: From Rabi Oscillations to Fourier Axis Controlling the Musical Timbre

Reiko Yamada [1]✉, Samuele Grandi [1], Gorka Muñoz-Gil [1], Luca Barbiero [1,3], Albert Aloy [1] and Maciej Lewenstein [1,4]✉

[1]ICFO - Institut de Ciencies Fotoniques, The Barcelona Institute of Science and Technology, Av. Carl Friedrich Gauss 3, 08860 Castelldefels (Barcelona), Spain
[2]Fundació Phonos, Universitat Pompeu Fabra, Roc Boronat 138, 08018, Barcelona, Spain
[3]Institute for Condensed Matter Physics and Complex Systems, DISAT, Politecnico di Torino, I-10129 Torino, Italy
[4]ICREA, Pg. Lluís Companys 23, 08010 Barcelona, Spain
[1]✉ Corresponding author: reiko.yamada@icfo.eu
[1]✉ Corresponding author: maciej.lewenstein@icfo.eu

## ABSTRACT

Randomness has attracted great interest in the field of music composition for quite some time. As early as 1962, Iannis Xenakis started exploring a stochastic approach to randomness by using computer-based interlinking probability functions to determine compositional structure, pitches and their durations [15]. Soon after, composers and music technologists started to explore randomness with various methods of algorithmic compositions, sometimes with the help of artificial intelligence. However, in most cases, the source of randomness they used was in fact deterministic in nature. That is to say, the random numbers that they employed are imperfect in the strict sense (simply put, perfect random numbers never have repeating patterns). Moreover, the method in which they produced such randomness was extrinsic to the method in which randomness was applied. In this project, we attempt to take a further step by directly producing sound events from the genuine quantum true randomness of quantum physical systems. Through this method, we aim at achieving a new sense of aesthetic effect in music which derives from the true randomness that prevails in the natural quantum world.

## 1   Introduction

Imperfection plays an important and often positive role in aesthetics. For example, the color of blue, yellow brown, green, purple, pink, orange and red diamonds is due to small numbers of defects or impurities in crystal lattice atoms [19]. Our research is first and foremost, to explore randomness, a concept related to imperfections in the domain of quantum mechanics, in order to generate inspiration and space for new approaches in art making, in particular in the field of computer-assisted and AI based music composition.

The field of quantum mechanics revolutionized modern science in the areas of sensing, communication, computing and cryptography by representing the world in a non-deterministic, chance-based model, in which intrinsic randomness is a core concept. In this paper, after defining the difference between classical (apparent) and quantum (intrinsic) randomness, we provide a quick view of adaptations of randomness in the history of modern music composition. In the following sections, using two musical excerpts, we demonstrate difficulty to distinguish the aesthetic effects of two kinds of randomness: apparent and intrinsic. We raise the question of the methods which have been widely employed in computer and AI assisted compositional processes to employ randomness. Finally, we present our original method of using quantum randomness observed in the certain concrete physical process, the so-called Rabi oscillation, in the creative process.

## 2    Imperfection and Randomness

The aesthetic concept of imperfection, under various names, has been explored in many different artistic contexts. For example, the Japanese traditional aesthetic concept of *Wabi-sabi* has been cherished for centuries, as an appreciation for imperfect, incomplete and transient beauty [26]. In 1913, composer Luigi Russolo constructed mechanical noise generators which he called *intonarumori* and introduced the concept of noise music with his manifesto *L'arte dei rumori*. In the 1990s, Glitch emerged as a sub-genre of electronica and popularized the use of noise in the general public. In concert music of the 20th and 21st century, and in free improvisation of recent years, many have explored extended techniques on traditional instruments, the use of non-traditional instruments and field-recorded sound materials such as *objet sonore*[1]. These examples show rising interest in incorporating sounds that are imperfect or not fully controlled.

This interest might stem from a willingness to express ourselves and our surrounding environment as the way it is, just like the world view of *Wabi-sabi* accepts imperfection, incompleteness and transience. Randomness is considered to be one of ramifications of the aesthetic concept of imperfection [23]. In generative computer art, "randomness is often used to humanize or introduce variation and imperfections to an underlying rigid deterministic process."[2] With the introduction of computer-assisted art making, one is able to precisely control the degree of perfection or predictability, for example, by adding an element of noise with the use of pseudo or true randomness. Perhaps, in order to further accept ourselves and our surroundings as the way it is through music making, one can seek a method in which the true randomness from the natural world is shown with minimum interference of the creator.

Randomness has received significant interest in both quantum mechanics and computer-assisted compositions for some time. In recent years, a number of scientific and artistic efforts have been made to connect the two fields using concepts such as randomness. When discussing randomness in physics, it is extremely important to distinguish the difference between apparent randomness and intrinsic randomness. "Apparent randomness is the randomness that results exclusively from a lack of full knowledge about the state of the system in consideration" [3]. Such an approach generally corresponds to the determinist perspective in classical physics. On the contrary, quantum physics posits that "Intrinsic randomness is the randomness that persists even if we have the full knowledge about the state of the system in consideration" [3]. According to this perspective, a given system can be fully or partially intrinsically random; in such a case, probabilities and stochastic processes are necessary to understand the system's environment and to predict its behaviour in the future.

## 3    Apparent Randomness

Composers of the 20th and 21st centuries who have experimented with the use of randomness most often used chance operations or random number generators in their creative process. While their motives for utilizing randomness have been extremely diverse, the randomness they employed has almost entirely been apparent randomness.

Looking at historical examples, John Cage, first attempted to use chance as a governing principle in his score for *Music of Changes* in 1951. By adapting I-ching, an ancient Chinese complex random number operation, Cage tried to find a way to bypass a reliance on his aesthetic judgement. While Cage used chance operations to organize his pre-composed materials in *Music of Changes*, Pierre Schaeffer started from concrete sounds and used them to build a structure with *musique concrète*. It was his commitment to compose with materials taken from 'given' experimental sound, sound fragments that are considered as discrete and complete sound objects as they are [24].

In recent years, algorithmic and stochastic computer-assisted composition and the use of Artificial Intelligence have taken center stage. One of the most notable composers to pave this way is Iannis Xenakis who, among other methods, used high-speed computer computations to calculate various probability theories to assist compositional processes. The program would "deduce" a score from a "list of note densities and probabilistic weights supplied by the programmer, leaving specific decisions to a random number generator" [2]. Yet in all of these cases, the form of randomness used was apparent (or pseudo) randomness, which is ultimately deterministic in nature. Although each outcome using different systems varies wildly, the difference that using true, intrinsic random numbers would create remains unexplored.

---

[1] (sound object) a term coined by Pierre Schaeffer
[2] Extended version of: McCormack, J., Bown, O., Dorin, A., McCabe, J., Monro, G., and Whitelaw, M., "Ten Questions Concerning Generative Computer Art", *Leonardo*, MIT Press, (2012): 10.

## 4    Intrinsic Randomness

How can we best incorporate intrinsic randomness in the creative process of music composition, and would musical effects be noticeably different from the outcome of mere apparent randomness? In fact, nowadays it is easily possible to use truly intrinsic random numbers, for example by using quantum random generated numbers that are streamed and available online[3]. However, in order to determine its musical effect, in comparison to using apparent random numbers, one must examine the method in which the random numbers are conventionally mapped to musical parameters.

To demonstrate this point, we start by presenting the three musical excerpts below. For this experiment, we mapped random numbers to the order of musical events that they appear. The sound events are re-organized using two different random numbers: intrinsic and apparent respectively. The first twenty sound events from the source material, free improvisation by Jordina Millà Benseny[4] (Fig. 1), have been reorganized using a set of quantum random numbers[5] (Fig. 2) and by a set of classical random numbers[6] (Fig. 3).

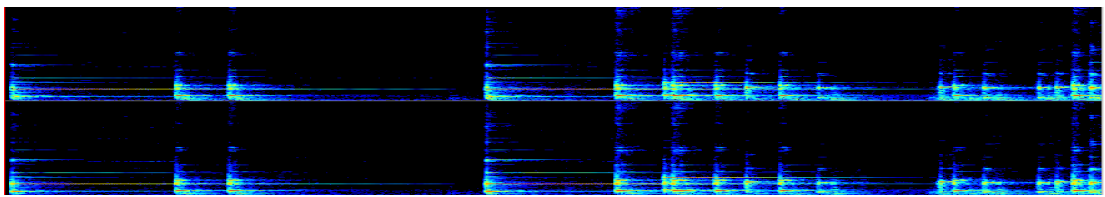

**Figure 1**. First twenty musical events from the free improvisation by Jordina Millà Benseny.

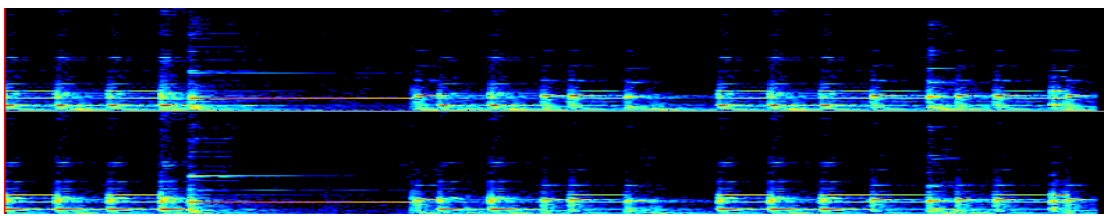

**Figure 2**. The excerpt reorganized by quantum random intrinsic numbers.

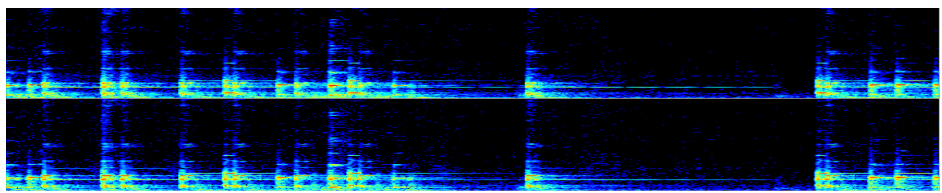

**Figure 3**. The excerpt reorganized by classical random apparent numbers.

---

[3] For example, https://qrng.anu.edu.au/
[4] Jordina Millà Benseny, first twenty sound events from *dianthus pirenaicus* the (track 8) of *Males Herbes*
[5] Generated on the patch written by an author using "random" object in Max/MSP/Jitter 8 (Cycling '74)
[6] Quantum random numbers provided by QUSIDE (www.quside.com)

In these short examples above, the difference in musical effect between the use of apparent and intrinsic randomness is too small to be humanly perceptible. This, however, is partly expected due to the short length of these musical events. It also raises the question of the quality of random numbers that has been used in this experiment, and how it affects the overall musical effect, as well as the type of sound parameters that are suitable for this experiment.

In order to test how perfect, the set random numbers is, physicists measure the so-called correlation length. The correlation length in random numbers is a span before a set of numbers starts or stops repeating themselves. In general, quantum random numbers have much longer correlation length than classical random numbers, or they do not repeat at all, which means that a set of random numbers is absolutely perfect. In order to observe the difference between classical and quantum randomness expressed as a row of numbers, an extremely large set of numbers is therefore required.

Likewise, if we were to translate into music the above examples in order to measure the degree of perfection of the randomness that guides them using this method, the musical events would need to be infinitely longer. Such extension in length would also not make the differences any more visible to the human audience, by nature limited in its ability to memorize and recognize patterns. For the above examples, we did not use enough numbers to reveal the quality of randomness; this method of measuring randomness therefore does not appear to be the most useful in this case.

For this reason, it seems preferable to consider an application of random numbers in the reciprocal axis, controlling, for instance, frequency Fourier spectrum of the sound parameters instead. The frequency content of sound determines timbre, the sound color. Therefore, we explored the possibility of mapping randomness onto timbral changes in sounds. In the quantum world, patterns emerge whenever there is an imbalance in the probability distribution. Consequently, using perfect randomness from quantum physics should prevent the appearance of any patterns in timbral changes, whereas using classical randomness would eventually reveal some

## 5 Derivation of Sound Parameters Directly from Natural Random Phenomena

Mapping randomness on the Fourier axis, however, requires questioning the way in which randomness (of any type) has been translated in music vocabulary, that is to say, by converting randomly generated numbers into discrete musical parameters. Is there a different way in which intrinsic randomness and its related phenomena can directly relate to sound phenomena? Could randomness be integrated into composition without needing to be translated into numbers and introduced in the music via pre-selected sound parameters? We propose a new method, in which certain physical random phenomena, such as Rabi oscillations in a multi-level quantum system, directly produce random transformation of timbre. Such direct mapping of a physical random phenomenon into sound parameters, potentially involving minimal human interference in the translation process, might reveal another kind of beauty in imperfection. Just like how we find beauty in crystals and their natural imperfections, we would like to hear the sound timbres expressing imperfections directly derived from the random phenomena happening in physical oscillators.

## 6 Methodology

One of the foundations of quantum mechanics is that quantum systems have discrete energy levels. In the case of atoms, these levels describe, roughly speaking, the electron-nucleus distance: they may have either electronic energy $E_1$ or energy $E_2$, with increasing distance from the nucleus, but nothing in between. This is why we call quantum physics as we do: nearly every physical quantity is quantized, attaining only discrete values. In this picture, the state of an atom can be described by its probability of being found in any of the available energy states. Without external energy sources, an atom is usually found in the state with lowest energy, which we call the *ground state*. When shining light onto an atom with a laser that has a specific energy equal to the separation between two states, i.e. $E_1$-$E_2$, the atom will start to move back and forth between these two levels following a periodical pattern, in a process that is called Rabi Oscillations [25]. Consequently, the probability of measuring the atom in a particular electronic state increases and decreases over time, as seen in Fig. 4, with a frequency related to the properties of the atom and the laser. Following the assigned probabilities, each time we measure in which state the atom is it will be randomly found in any of the possible states. Remarkably, the randomness underlying this procedure is an intrinsic property of quantum nature. In other words, we are in the presence of *true* randomness.

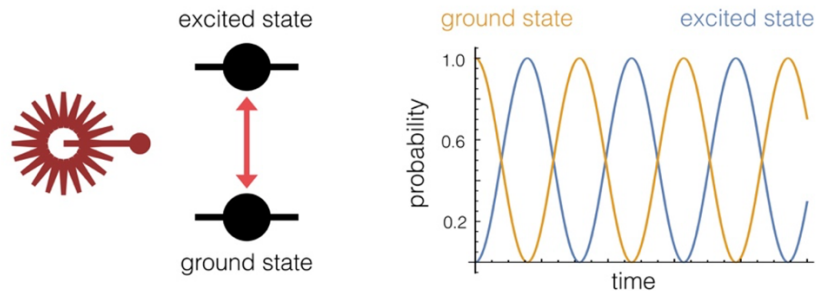

**Figure 4**. Rabi oscillation in a two-level system at resonance condition.

We match these sinusoidal Rabi oscillations with sound waves containing complex upper partials (harmonics). While the fundamentals of the oscillators always remain intact, the upper partials can be divided into several groups of timbres. Depending on the state of the atom when measured, the fundamental is combined with one group of upper partials or another, creating sound color to change. The probability of occurrence in any given timbre is directly determined by the probability of the atom to be in its different possible states, directly related to the Rabi oscillations as stated above.

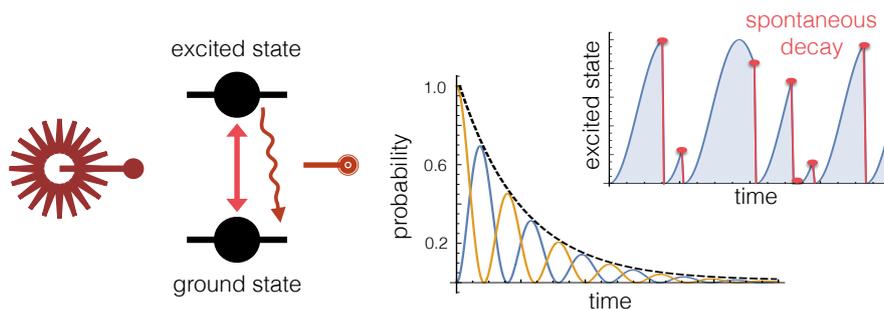

**Figure 5.** Rabi oscillation and spontaneous emission. The inset shows the dynamic of the excited state population.

Another interesting intrinsically random process related to this topic is the so-called spontaneous emission. When an atom is excited from the ground state to another one with a higher energy, that is called *excited state*, it can spontaneously decay back to the ground state shining the "excess" of energy in the form of light (and more precisely, photons). This event occurs at random times, with a distribution following an exponential decay. When mixing the effects of spontaneous emission and Rabi oscillations, the result is simple: the probabilities continue oscillating as they were; then, when the spontaneous emission occurs, the atom falls to the lower energy state, meaning that the probability of finding the atom in this state is one (that is, 100%); finally, the probabilities start oscillating again from this point, with similar frequency as before this event. The new probability for the atom to be in the excited state is shown in Fig. 5. Whenever the atom decays from the excited state to the ground state a photon is emitted, which can be detected by suitable detectors. Again, the randomness involved in this process stems from the intrinsic random nature of quantum mechanics and is not anyhow related to an incomplete knowledge of such a system. Note that the rate at which spontaneous emission occurs changes vastly from one system to another, from nanoseconds to even days in the case of very stable atoms or molecules. The frequency of the Rabi oscillation depends on the physical properties of the atom, as well as from the power of the excitation laser light.

We combine the randomness of the spontaneous decay of the atom with the associated probability of finding it in the excited state. As visible from the inset in Fig. 5, the atom oscillates between the excited and the ground state. However, since the spontaneous decay only happens from a higher to a lower energy state, these events will be more likely to occur when the atom has a higher probability of being in the excited state. Consequently, by collecting the times at which the atom decays, the resulting histogram would be identical to the probability shown in Fig. 5, for a high number of collected events. However, the intermediate histograms, and especially the first one, would be rather different from the final result. If we turn the axis of time into frequency, representing for example higher harmonics of a fundamental note, and the height of the histogram as intensities, then we could play these harmonics in succession. As spontaneous emission occurs at random times, the time intervals before every timbral change is bound to be truly random and obeying only the combined probability distributions of spontaneous emission and Rabi oscillation.

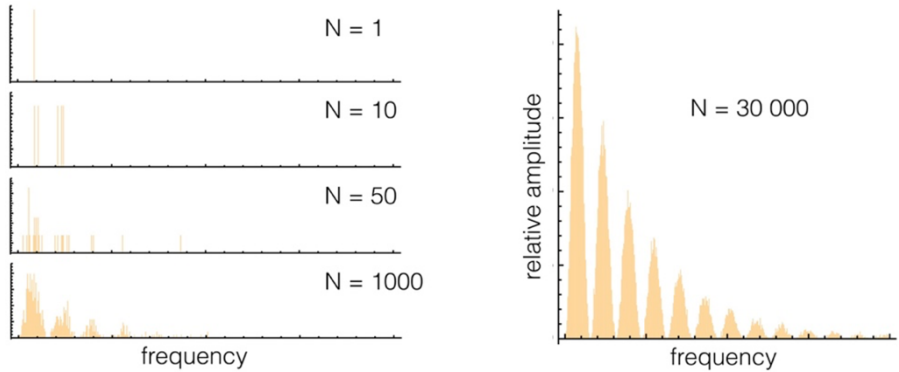

**Figure 6**. Example of histograms of times between consecutive spontaneous emissions, for increasing numbers of simulated events.

The result would be a sound going through changes at random times, by the random introduction of an extra harmonic, all according to the probability distribution of spontaneous emission and Rabi oscillations. Such a progression is shown in Fig. 6.

One of the most original features in our method is that the random events within a certain time frame are represented as one palette of musical timbres (Fig. 7) within one musical event. This method allows the comparison of multiple samples of randomness to be listened and analyzed in a short amount of time. As a result, we do not convert certain random phenomena into numbers and then convert them again into sound parameters. In contrast, the intrinsic random phenomena here is directly seen and heard through timbres one by one, a few seconds apart.

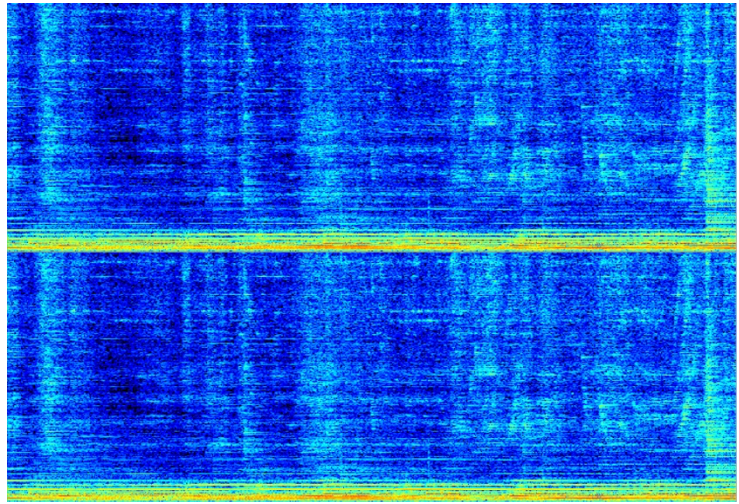

**Figure 7**. Example of sonogram analysis of a timbre containing multiple fundamentals with rich upper partials.

## 7    Conclusion and Future Work

After we reviewed the general use of randomness in computer-assisted compositional methods, we drew the primary conclusion that apparent randomness of quantum origin might be better utilized if placed in the reciprocal axis of the sound parameters. We propose to apply quantum randomness to the frequency spectrum of the sound parameters, therefore the random state of the atom to control the timbre of the sounds. The resulting sound, consisting of multiple fundamentals and rich upper partials, is directly derived from the quantum true randomness with minimal human interference in a translation process.

Essentially, we are interested to see whether true randomness can provide us a renewed aesthetic effect if used with the method and the contexts directly derived from the source of the randomness, however subtle it could be. The comparison between an image from the structure of hexagonal boron nitride, and a Penrose tiling may be used as an analogy. The perfect structure of hexagonal boron nitride from nature, and a human made Penrose tiling (which intentionally contains imperfect patterns) may be similar at first sight, however their aesthetic effects are profoundly different.

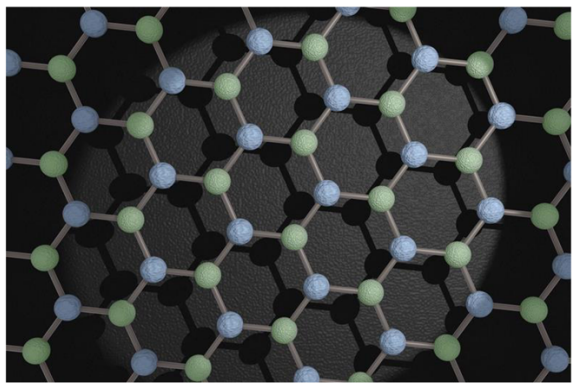

**Figure 8**. Hexagonal boron nitride Source[7].

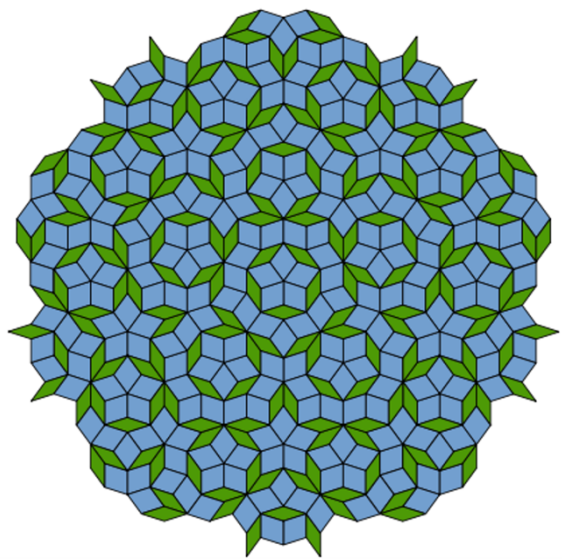

**Figure 9**. Penrose tiling using thick and thin rhomb.

Some quantum phenomena are still mysterious, and difficult to grasp rationally or intuitively. Timbre is one of the most abstract elements in sounds, creating a number of aesthetic effects that we do not yet have tools to analyze. We are at the beginning stage of an experiment that still requires much fine-tuning and trials in both of its sound synthesis and quantum physics components. After multiple trials of our proposed method, we hope that the mysterious quantum effect, with the help of quantum randomness, will reveal itself through sound colors, and in turn open the door for new aesthetics and new tools in computer-assisted music creation.

---

[7] The image taken from: Science Photo Library/Getty Images

## Acknowledgements

We thank Dr. Antoine Reserbat-Plantey (ICFO), Dr. Brian Cherney (McGill University) and Dr. Nicolas Trepanier (University of Mississippi) for valuable discussions and suggestions on the manuscript. ICFO group acknowledges support from ERC AdG NOQIA, State Research Agency AEI (Severo Ochoa Center of Excellence CEX2019-000910-S, Plan National FIDEUA PID2019-106901GB-I00/10.13039 / 501100011033, FPI, QUANTERA MAQS PCI2019-111828-2 / 10.13039/501100011033), Fundació Privada Cellex, Fundació Mir-Puig, Generalitat de Catalunya (AGAUR Grant No. 2017 SGR 1341, CERCA program, QuantumCAT / U16-011424, co-funded by ERDF Operational Program of Catalonia 2014-2020), EU Horizon 2020 FET-OPEN OPTOLogic (Grant No 899794), and the National Science Centre, Poland (Symfonia Grant No. 2016/20/W/ST4/00314), Marie Sklodowska-Curie grant STREDCH No.101029393, La Caixa Junior Leaders fellowships (ID100010434), and EU Horizon 2020 under Marie Sklodowska-Curie grant agreement No.847648 (LCF/BQ/PI19/11690013, LCF/BQ/PI20/11760031, LCF/BQ/PR20/11770012).